Research

Artificial Intelligence—Review

# Advancing Financial Engineering with Foundation Models: Progress, Applications, and Challenges

Liyuan Chen [a,b], Shuoling Liu [b], Jiangpeng Yan [a,b], Xiaoyu Wang [b], Henglin Liu [a], Chuang Li [a], Kecheng Jiao [a], Jixuan Ying [a], Yang Veronica Liu [c], Qiang Yang [c,*], Xiu Li [a,*]

[a] *Tsinghua Shenzhen International Graduate School, Tsinghua University, Shenzhen 518055, China*

[b] *E Fund Management Co., Ltd., Hengqin 519031, China*

[c] *Department of Data Science and Artificial Intelligence, The Hong Kong Polytechnic University, Hong Kong 999077, China*

*Corresponding authors.

*E-mail addresses*: profqiang.yang@polyu.edu.hk (Q. Yang), li.xiu@sz.tsinghua.edu.cn (X. Li)

**ABSTRACT**

The advent of foundation models (FMs), large-scale pre-trained models with strong generalization capabilities, has opened new frontiers for financial engineering. While general-purpose FMs such as GPT-4 and Gemini have demonstrated promising performance in tasks ranging from financial report summarization to sentiment-aware forecasting, many financial applications remain constrained by unique domain requirements such as multimodal reasoning, regulatory compliance, and data privacy. These challenges have spurred the emergence of financial foundation models (FFMs): a new class of models explicitly designed for finance. This survey presents a comprehensive overview of FFMs, with a taxonomy spanning three key modalities: financial language foundation models (FinLFMs), financial time-series foundation models (FinTSFMs), and financial visual-language foundation models (FinVLFMs). We review their architectures, training methodologies, datasets, and real-world applications. Furthermore, we identify critical challenges in data availability, algorithmic scalability, and infrastructure constraints and offer insights into future research opportunities. We hope this survey can serve as both a comprehensive reference for understanding FFMs and a practical roadmap for future innovation.

## 1. Introduction

Financial engineering, an interdisciplinary domain that combines finance, mathematics, and computer science, has become the cornerstone of modern financial systems. It plays a critical role in designing complex financial products, managing risk, and supporting decision-making through quantitative models and data-driven algorithms [1]. As financial markets grow in complexity and data heterogeneity, there is an increasing need for intelligent systems [2–5] that can generalize across a wide range of tasks, adapt to changing environments, and support scalable, automated financial workflows.

Recent advancements in foundation models (FMs)—that is, large-scale pre-trained models such as GPT-4 [6], Gemini [7], Qwen [8], and DeepSeek [9]—have demonstrated impressive generalization capabilities across modalities and domains. In financial contexts, these general-purpose FMs have already begun to reshape traditional pipelines, offering strong performance on tasks such as financial report summarization, portfolio commentary generation, sentiment-aware forecasting, and risk disclosure analysis [10–12]. This emerging paradigm shift from narrowly specialized artificial intelligence (AI) systems to general-purpose FMs promises to transform the landscape of financial engineering, as illustrated in Fig. 1.

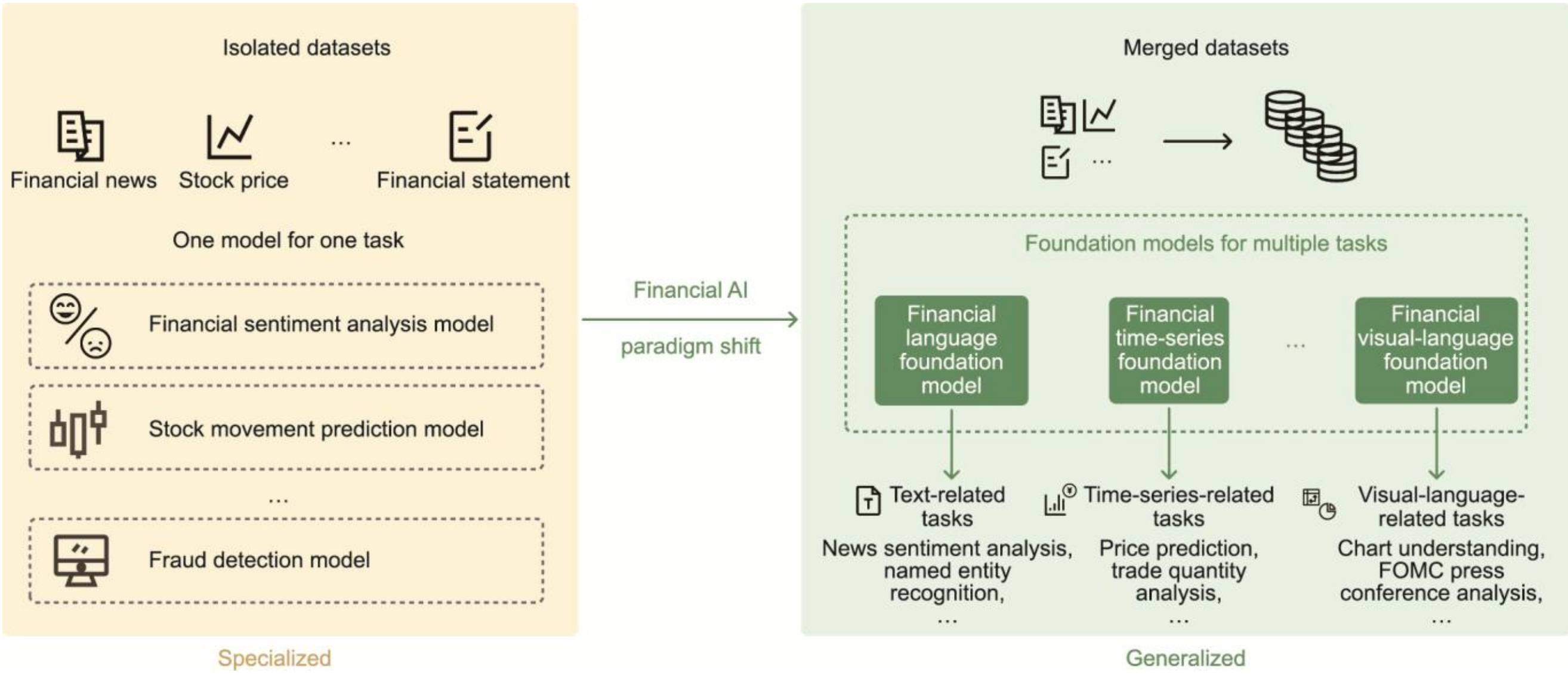

**Fig. 1.** The financial AI paradigm is shifting from the use of traditional models for specific tasks to the use of FMs for multiple tasks. FOMC: Federal Open Market Committee.

Despite the strong potential of general-purpose FMs in finance, many financial tasks possess unique characteristics, such as the need for legal compliance, multimodal document processing, long-horizon time-series analysis, and strict privacy requirements, that are not fully addressed by off-the-shelf models. These challenges have motivated the research of financial foundation models (FFMs)—a class of FMs designed specifically for financial scenarios through domain-aware pre-training, task-specific fine-tuning, and alignment with financial reasoning and regulatory goals.

In this survey, our aim is to systematize the development of FFMs. We categorize FFMs into three major types based on input modality and application scope:

- **Financial language foundation models (FinLFMs).** Also referred to as financial large language models (FinLLMs) [13,14], FinLFMs are models pre-trained on financial texts, including reports, news, and contracts, and optimized for tasks such as question answering, summarization, and compliance checking [15,16].
- **Financial time-series foundation models (FinTSFMs).** FinTSFMs extend FFMs beyond language by processing sequential financial data (e.g., price histories and economic indicators) for financial time-series data analysis. Unlike texts, financial time series are continuous, non-stationary, and often multivariate, requiring inductive biases for temporal dependencies, regime shifts, and volatility clustering [17].
- **Financial visual-language foundation models (FinVLFMs).** FinVLFMs are designed to process both textual and visual information, such as financial charts, tables, and figures, which enables them to perform complex tasks that require multimodal understanding. These models address the growing need for cross-modal reasoning, in which financial knowledge is communicated through both textual narratives and visual artifacts [18–20].

These three categories collectively represent the ongoing shift from narrow task-specific modeling to broadly adaptable and general-purpose systems, which is enabling a new era of scalable, multimodal, and intelligent financial engineering [10,16,21]. In this survey, we aim to explore the current state of FFMs comprehensively. The following questions guide our study: ① Given that FMs have made remarkable achievements in other fields, what is the current progress of FFMs in the financial industry? ② With the development of these models, what challenges do they encounter in terms of data, algorithms, and infrastructure? By answering these questions, we hope to provide a comprehensive overview of the current situation of FFMs and offer a clear vision for their future development. In this article, we focus on the current progress of FinLFMs, FinTSFMs, and FinVLFMs from 2018 (the start of the FM era in a broader sense [13]) to June 2025, along with associated challenges and opportunities. We believe this survey will assist financial researchers and practitioners in quickly grasping the development of FFMs and will inspire new ideas for further innovation.

### *1.1. A brief history of FMs in finance*

The evolution of FFMs mirrors key milestones in the broader development of AI. The launch of bidirectional encoder representations from transformers (BERT) in 2019 [22] marked the start of the pre-training era, leading to early finance-specific BERT-style varients (FinBERT) [23–25] tailored for financial text understanding. This initiated the first wave of FinLFMs. The emergence of GPT-3 and ChatGPT, built on the now-famous scaling law [26], demonstrated the power of massive autoregressive models and instruction tuning. These advances inspired a new generation of FinLFMs that leveraged generative capabilities, human feedback alignment, and multi-task transfer [13,14]. However, early FinLFMs trained solely on text data perform poorly when directly applied to numerical time series, as they lack inductive biases for sequential dynamics such as seasonality, volatility, and cross-asset dependencies. This limitation motivated the development of FinTSFMs [27], which either pre-train directly on historical price data or adapt large language models (LLMs) with temporal encoding modules, highlighting the distinct requirements of financial time-series forecasting.

As a further challenge, financial information is not only embedded in text and raw sequences but also conveyed through visual modalities such as the interest rate “dot plot” and projection charts at the Federal Open Market Committee (FOMC) press meetings, tabular statements, and regulatory filings. The advent of multimodal FMs such as GPT-4o [28] inspired the design of FinVLFMs [18–20], which align visual and textual streams to support tasks such as chart-based question-answering (QA), table interpretation, and report figure analysis. This line of research emphasizes that FinVLFMs are not substitutes for FinTSFMs but rather complementary tools that capture the visual-symbolic layer of financial communication.

Together, the progression from FinLFMs to FinTSFMs and FinVLFMs marks the trajectory of FFMs toward broader generality, stronger reasoning, and richer modality integration in financial AI.

### *1.2. Comparison of related surveys with our contributions*

In our extensive literature search, we identified nine representative surveys on FFMs. These existing works have provided valuable insights from various perspectives. However, compared with these prior studies, our survey offers a more comprehensive and in-depth examination of FFMs, spanning methodologies, data sources, and applications, along with detailed discussions on current challenges and future research directions. Table 1 [10–12,16,21,29–32] summarizes the key distinctions between our survey and the existing literature. Specifically, our survey presents the following unique contributions:

**(1) A systematic taxonomy and thorough coverage of FFM sub-fields.** Our survey categorizes FFMs into three major sub-fields—language, time-series, and visual-language models—in the context of financial tasks. This comprehensive taxonomy provides a unified view of the entire FFM landscape, surpassing the scope of existing surveys that tend to focus on specific modalities.

**(2) A holistic and deeper exploration of multiple dimensions of FFMs.** We investigate FFMs from a range of perspectives, including modeling techniques, data characteristics, practical applications, technical challenges, and potential future directions. Compared with those in previous reviews, this integrated approach allows readers to gain a more complete and nuanced understanding of FFMs.

**(3) An up-to-date and practitioner-oriented synthesis of methods and resources.** Beyond reviewing published work, we provide a curated and continuously updated repository of FFM-related papers, tools, and datasets on GitHub entitled “finllm/Awesome-FinFMs”. This living resource serves as a practical reference hub for researchers and practitioners, something that is rarely addressed in other surveys.

**Table 1**

Comparison between our survey and existing surveys on FFMs.

| Reference | FinLFMs | FinTSFMs | FinVLFMs | Datasets | Applications | Challenges |
|---|---|---|---|---|---|---|
| [21] | ○ | × | × | × | ○ | ○ |
| [29] | × | × | × | × | ✓ | ○ |
| [16] | × | × | × | ✓ | ✓ | ○ |
| [10] | ✓ | × | × | ✓ | ✓ | ✓ |
| [11] | ○ | × | × | ○ | ○ | ○ |
| [12] | × | × | × | ✓ | ○ | ○ |
| [30] | × | × | × | × | ○ | ○ |
| [31] | × | × | × | × | ✓ | ✓ |
| [32] | ○ | × | × | ○ | ○ | ○ |
| Our survey | ✓ | ✓ | ✓ | ✓ | ✓ | ✓ |

× indicates no content discussion, ○ indicates brief and non-comprehensive discussion, and ✓ indicates comprehensive discussion.

### *1.3. Contributions*

In this survey, we contribute to the field by providing the following:

**(1) A systematic review of methods (Sections 2–4).** We propose a novel taxonomy of FFMs and conduct a comprehensive review of existing methods across the three major sub-fields: language (Section 2), time-series (Section 3), and visual-language (Section 4). We also investigate different training strategies, offering insights into potential technical innovations for FFMs.

**(2) A comprehensive survey of datasets (Sections 2–4).** Alongside the model review, we survey a wide range of large-scale datasets and databases that are applicable to training FFMs in different modalities. This survey helps identify key limitations of current financial data resources and provides practical guidance for dataset selection in future research.

**(3) A thorough overview of applications (Section 5).** We present an overview of diverse financial applications enabled by FFMs, illustrating how these models are currently being deployed in real-world financial engineering tasks. This serves as a reference point for extending FFMs to new application scenarios.

**(4) An in-depth discussion of key challenges and future research (Section 6).** We analyze critical challenges related to data, algorithm, and computing infrastructure in the development and deployment of FFMs. These discussions identify major bottlenecks and open new directions for future research.

## 2. Financial language foundation models

To start, we introduce FinLFMs in this section. FinLFMs are a class of domain-specific language foundation models (LFMs) that are specifically trained on financial corpora to acquire text comprehension, reasoning, and generation capabilities tailored for financial scenarios. We investigate FinLFMs through three analytical lenses:

**(1) Collection and category.** This lens focuses on how the reviewed FinLFMs are collected and classified into BERT-style FinLFMs, generative pre-trained transformer (GPT)-style FinLFMs, and reasoning-enhanced FinLFMs.

**(2) Training method analysis.** Through this lens, we investigate innovations across pre-training strategies, supervised fine-tuning techniques, and alignment mechanisms for training FinLFMs.

**(3) Language-related datasets.** Here, we introduce commonly used datasets in the training and evaluation of FinLFMs.

### *2.1. Model classification*

To comprehensively review FinLFMs, we conducted an extensive search across multiple channels (e.g., Google Scholar, Web of Science, and arXiv) and scoured GitHub repositories and official announcements from major financial institutions and tech companies for open-source and proprietary FinLFM projects. In total, our survey covers 21 representative FinLFM models.

The evolution of FinLFMs closely mirrors that of general LFMs. Fig. 2 shows the development trajectory of the reviewed FinLFMs based on their architectural backbones, with a timeline. Guided by this timeline, we categorize the FinLFMs into three groups: BERT-style, GPT-style, and reasoning-enhanced FinLFMs.

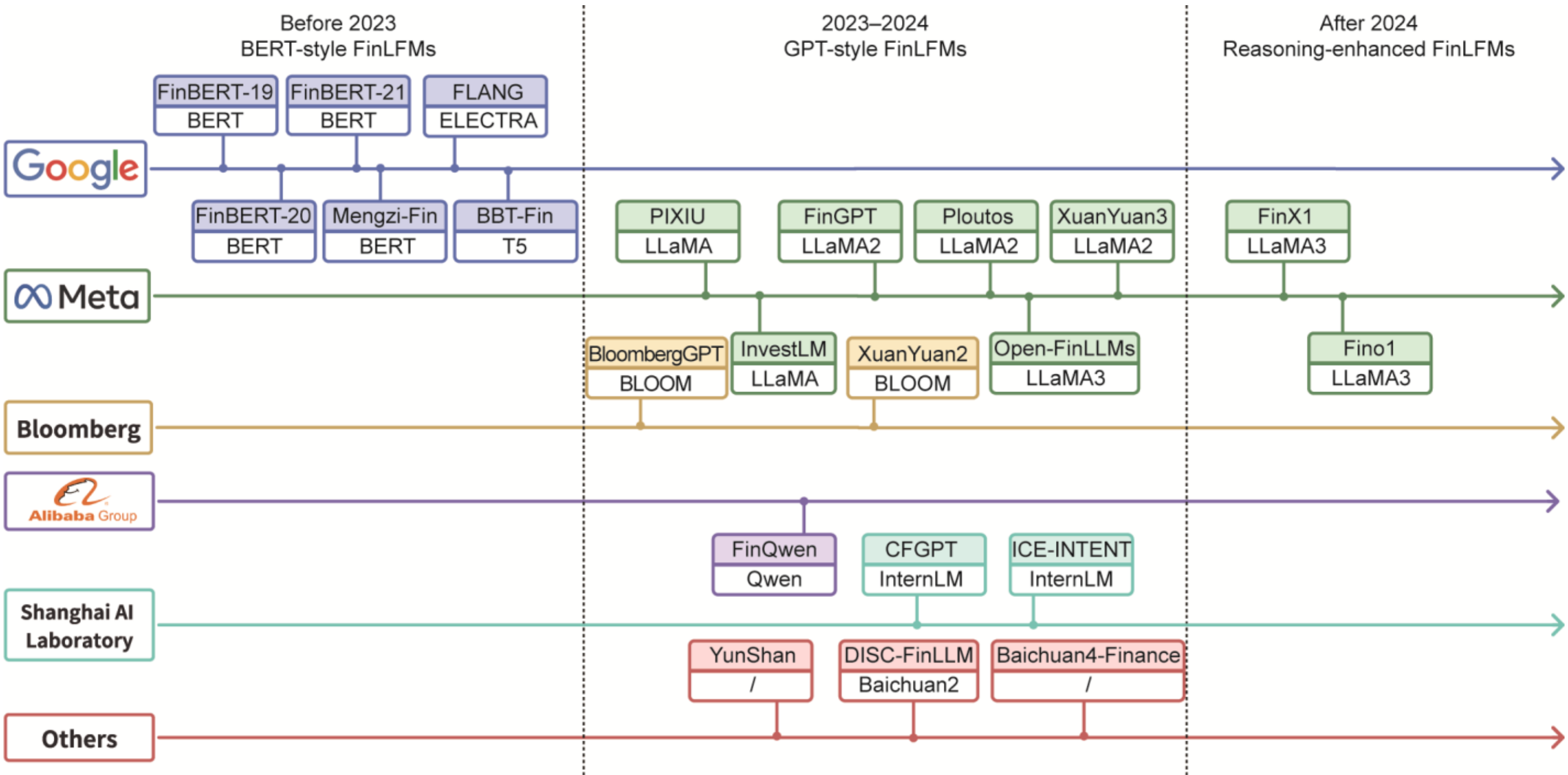

**Fig. 2.** The development trajectory of FinLFMs based on their architectural backbones, shown along a timeline. Each horizontal strip corresponds to an institution that released a backbone FM; the tiles arrayed along the strip denote FinLFMs developed atop one of that institution's backbones. The name of each FinLFM appears in the upper portion of the tile, while its corresponding backbone model is listed in the lower portion. T5: text-to-text transfer transformer; ELECTRA: efficiently learning an encoder that classifies token replacements accurately; FinBERT-19/20/21: financial-specific BERT-style variant developed in 2019/2020/2021; Mengzi-Fin: a BERT-style financial model trained for Chinese texts; BLOOM: BigScience large open-science open-access multilingual language model; FLANG: financial LANGuage model; LLaMA: a family of LLMs developed by Meta; CFGPT: Chinese financial GPT; BBT-Fin: BigBang transformer financial model; DISC-FinLLM: data intelligence and social computing laboratory-financial LLM; InternLM: a family of LLMs developed by Shanghai AI Laboratory; Fino1: an open financial reasoning model trained via supervised fine-tuning and group relative policy optimization (GRPO)-based reinforcement learning; Open-FinLLMs: open LLMs for financial applications;

InvestLM: an LLM for investment using financial domain instruction tuning; Ploutos: an interpretable financial LLM for stock movement prediction; XuanYuan2: a large Chinese financial chat model from Du Xiaoman Technology; FinGPT: an open-source financial LLM from AI4Finance Foundation; PIXIU: a financial LLM from ChanceFocus; Qwen: a family of LLMs developed by Alibaba Cloud; FinQwen: an LLM for financial domain based on Qwen base model; ICE-INTENT: a Chinese–English bilingual financial by fine-tuning the InternLM backbone model; YunShan: a specialized LLM for domains of finance and law from Huawei; Baichuan4-Finance: an aligned language model tailored for finance domain from Baichuan Intelligent Technology; Baichuan2: a series of large-scale multilingual language models from Baichuan Intelligent Technology; FinX1: a financial industry reasoning big model from Du Xiaoman Technology.

#### *2.1.1. BERT-style FinLFMs*

We use the term "BERT-style FinLFM" to refer to transformer language models that are primarily designed for bidirectional representation learning and discriminative tasks, typically trained with masked language modeling (MLM) or related objectives, inspired by the original BERT [22] architecture. Early efforts in this line produced the first finance-specific LFM variants. For instance, a financial-specific BERT-style variant developed in 2019 (FinBERT-19) [23] continued pre-training BERT on financial news, while another financial-specific BERT-style variant developed in 2020 (FinBERT-20) [24] and financial-specific BERT-style variant developed in 2021 (FinBERT-21) [25] were further pre-trained on different and mixed general and financial corpora. Additionally, Zhang et al. [33] explored the development of a BERT-based financial model trained for Chinese texts named Mengzi-Fin. Another example is financial LANGuage model (FLANG) [34], a finance-specialized variant of efficiently learning an encoder that classifies token replacements accurately (ELECTRA) [35], that uses selective masking tailored to financial keywords and phrases for better performance. More recently, with BigBang transformer financial model (BBT-Fin) [36], researchers introduced a text-to-text transfer transformer (T5)-style [37] financial pre-trained language model with knowledge-enhanced pre-training. Together, these BERT-style FinLFMs learn domain-specific vocabularies and patterns, forming a backbone for many later systems.

#### *2.1.2. GPT-style FinLFMs*

GPT-style FinLFMs are transformer language models trained with causal language modeling (CLM) objectives; they excel at autoregressive text generation and instruction following. These models inherit the generative strengths of GPT architectures and are increasingly specialized for financial applications through domain-specific pre-training or supervised fine-tuning. The rise of GPT-style FinLFMs was catalyzed by the success of ChatGPT [38], with BloombergGPT [15] marking a pivotal start. Built on the BigScience large open-science open-access multilingual language model (BLOOM) [39] architecture and trained on Bloomberg's proprietary data, BloombergGPT demonstrated the feasibility of domain-specific LLMs at scale. Since then, numerous finance-oriented variants have emerged, including a large Chinese financial chat model from Du Xiaoman Technology (XuanYuan2) [40] for Chinese markets. Crucially, the rise of GPT-style FinLFMs has paralleled the rapid evolution of their open-source foundations. As general-purpose LLMs such as Meta's LLaMA [41,42], Alibaba Cloud's Qwen [8], and Shanghai AI Laboratory's InternLM [43] continue to scale, FinLFMs have inherited architectural advancements such as deeper attention, longer context, and stronger instruction following, while layering in financial expertise via supervised fine-tuning or continual pre-training. This symbiotic relationship enables fast capability upgrades without training from scratch. For instance, Meta's LLaMA series has been used to underpin models such as a financial LLM from ChanceFocus (PIXIU) [44], an LLM for investment using financial domain instruction tuning (InvestLM) [45], an open-source financial LLM from AI4Finance Foundation (FinGPT) [46,47], open-source LLMs for financial applications (Open-FinLLMs) [19], an interpretable financial LLM for stock movement prediction (Ploutos) [48], and a large Chinese financial chat model from Du Xiaoman Technology (XuanYuan3) [49], while Qwen inspired an LLM for financial domain based on Qwen base model (FinQwen) [50]. InternLM [43] powers large-scale models such as a Chinese–English bilingual financial by fine-tuning the InternLM backbone model (ICE-INTENT) [51] and the Chinese financial GPT (CFGPT) [52], which combines extensive instruction datasets with financial domain knowledge. Collectively, these developments highlight how FinLFMs increasingly benefit from and contribute to the open-source LLM ecosystem. Commercial models also exist, such as a specialized LLM for domains of finance and law from Huawei (YunShan) [53] and an aligned language model tailored for finance domain from Baichuan Intelligent Technology (Baichuan4-Finance) [54], which follow similar patterns without disclosing full architectural details.

#### *2.1.3. Reasoning-enhanced FinLFMs*

More recently, a new class of FinLFMs has emerged with advanced reasoning and decision-making capabilities, inspired by models such as OpenAI-o1 and DeepSeek-R1 [55]. These systems leverage reinforcement learning to elicit long chain-of-thought (CoT) reasoning, enabling FinLFMs to generate multi-step analytical traces that improve factual accuracy and robustness in answering complex financial questions. For instance, a financial industry reasoning big model from Du Xiaoman Technology (FinX1) [49] and an open financial reasoning model trained via supervised fine-tuning and group relative policy optimization (GRPO)-based reinforcement learning (Fino1) [56] use CoTs to guide the models to generate intermediate reasoning steps before arriving at the final answer. These models represent a transition toward autonomous financial agents, built not only to respond but to reason, plan, and act across complex tasks with step-by-step CoTs.

### 2.2. Training method analysis

This section focuses on GPT-style FinLFMs, which now dominate the field in both research and deployment. Their training strategies are summarized in Table 2 [15,19,40,44,45,47–54,56,57], following the standard three-stage framework of pre-training, supervised fine-tuning, and alignment proposed in the *State of GPT 2023* report by Karpathy [58].

**Table 2**

Training details of representative FinLFMs.

| Model | Backbone | Method | Parameters | Additional training data | Reference | Open source |
|---|---|---|---|---|---|---|
| BloombergGPT | BLOOM | PT and SFT | 50 billion | [G] 345billion + [F] 363 billion tokens | [15] | — |
| PIXIU | LLaMA | SFT | 7 billion and 30 billion | [F] 136 000 instructions | [44] | https://github.com/The-FinAI/PIXIU |
| InvestLM | LLaMA | SFT | 65 billion | [F] 1 335 instructions | [45] | https://github.com/AbaciNLP/InvestLM |
| XuanYuan2 | BLOOM | PT and SFT | 176 billion | — | [40] | https://github.com/Duxiaoman-DI/XuanYuan |
| YunShan | — | PT and SFT | 7 billion | [F] 36.5 billion tokens | [53] | — |
| FinQwen | Qwen | PT and SFT | 14 billion | [F] 200 billion tokens | [50] | https://github.com/Tongyi-EconML/FinQwen |
| FinGPT | LLaMA2 | SFT | 7 billion | [F] 151 000 instructions | [47] | https://github.com/AI4Finance-Foundation/FinGPT |
| Ploutos | LLaMA2 | SFT | 7 billion | [F] ~30 000 instructions | [48] | — |
| DISC-FinLLM | Baichuan2 | SFT | — | [F] 250 000 instructions | [57] | https://github.com/FudanDISC/DISC-FinLLM |
| CFGPT | InternLM | PT and SFT | 7 billion | [F] 141 billion tokens + 1.5 million instructions | [52] | https://github.com/TongjiFinLab/CFGPT |
| ICE-INTENT | InternLM | SFT | 7 billion | [F] 604 000 bilingual samples from 36 datasets | [51] | https://github.com/YY0649/ICE-PIXIU |
| Open-FinLLMs | LLaMA3 | PT and SFT | 8 billion | [G] 18 billion + [F] 52 billion tokens | [19] | https://anonymous.4open.science/r/PIXIU2-0D70/B1D7/LICENSE |
| XuanYuan3 | LLaMA3 | PT, SFT, and alignment | 70 billion | ([G] + [F]) 275 billion tokens | [49] | https://github.com/Duxiaoman-DI/XuanYuan |
| Baichuan4-Finance | — | PT, SFT, and alignment | — | [G] 400 billion + [F] 100 billion tokens | [54] | — |
| FinX1 | LLaMA3 | SFT and alignment | 70 billion | — | [49] | https://github.com/Duxiaoman-DI/XuanYuan |
| Fino1 | LLaMA3 | SFT and alignment | 8 billion | [F] 5 000 instructions | [56] | https://github.com/The-FinAI/Fino1 |

PT: pre-training; SFT: supervised fine-tuning; [G]: general-purpose data; [F]: financial-specific data.

#### 2.2.1. Pre-training

Pre-training establishes the core language understanding and reasoning capabilities of FinLFMs. While early models such as BloombergGPT [15] were trained from scratch on massive financial corpora, recent trends favor continuous pre-training (CPT) on top of open-source general-purpose LLMs (e.g., LLaMA and Qwen), combining general and financial domains. This strategy, employed by models such as FinQwen [50] and Open-FinLLMs [19], enhances domain specialization without sacrificing generalization. Pre-training typically involves more than 100 billion tokens and requires a large-scale distributed infrastructure; thus, it remains mostly within the scope of industry laboratories. For instance, Baichuan4-Finance [54] employs 500 billion tokens across the general and financial domains. In contrast, academic models often skip this stage due to resource limitations, relying instead on downstream fine-tuning, which is introduced next.

#### 2.2.2. Supervised fine-tuning

Supervised fine-tuning (SFT) adapts a pre-trained model to financial downstream tasks using high-quality instruction datasets. A distinguishing feature of FinLFMs is the emphasis on instruction-following formats tailored to financial contexts, such as forecasting, question answering, risk assessment, and summarization. Recent models such as FinGPT [46,47], PIXIU [44], and Ploutos [48] leverage instruction datasets ranging from 30 000 to 150 000 examples. High-quality curation is critical: InvestLM [45], for example, demonstrates satisfying performance with only 1335 expert-crafted instructions. CFGPT [52] represents one of the most expansive efforts, using over 1.5 million financial instructions in addition to 141 billion tokens during pre-training. ICE-INTENT [51] also demonstrates the potential of multilingual and multi-dataset fusion by integrating over 600 000 bilingual financial instructions.

#### 2.2.3. Alignment

Alignment constitutes the final stage, in which FinLFMs are optimized to meet domain-specific behavioral objectives: factual accuracy, regulatory consistency, and reasoning transparency. Two major trends have emerged in alignment for financial models: ① Compliance-focused alignment, as seen in XuanYuan3 [49] and Baichuan4-Finance [54], employs reward models and reinforcement learning (e.g., proximal policy optimization (PPO)) to penalize hallucinations and ensure output consistency with legal and financial norms. ② Reasoning-centric alignment, exemplified by FinX1 and Fino1 [56], uses techniques such as GRPO to encourage detailed, interpretable CoT reasoning for high-stakes tasks. Despite its relatively nascent development, alignment is poised to become a cornerstone of future FinLFM research. As financial applications demand increasing levels of trustworthiness, interpretability, and regulatory compliance, alignment techniques offer a promising pathway for bridging model capabilities with domain-specific behavioral expectations. Future efforts will likely focus on developing standardized evaluation benchmarks, more robust reward modeling frameworks, and scalable alignment pipelines tailored to the financial domain.

### *2.3. Financial text-based datasets*

Table 3 [34,44,51,59–75] summarizes representative financial text-based datasets that have been developed to support the training and evaluation of FinLFMs. These datasets span a variety of tasks, languages, and modalities, reflecting the growing maturity and diversity of financial natural language-processing (FinNLP) resources. We categorize the evolution of financial-text datasets into three stages based on their scale, scope, and language coverage.

**Table 3**

Details of common financial text-based datasets.

| Dataset | Task(s) | Scale | Language | Open source |
|---|---|---|---|---|
| FPB [59] | Sentiment analysis | 4 840 financial news sentences | English | https://huggingface.co/datasets/takala/financial_phrasebank |
| NER [73] | Named entity recognition | 1 473 annotated sentences from 8 financial agreements | English | http://people.eng.unimelb.edu.au/tbaldwin/resources/finance-sec/ |
| FiQA [60] | Sentiment analysis and question answering | 675 microblogs + 498 headlines | English | https://sites.google.com/view/fiqa/home |
| ACL18 [61] | Stock prediction | Tweets and daily prices for 88 stocks (2014–2016) | English | https://github.com/yumoxu/stocknet-dataset |
| CIKM18 [62] | Stock prediction | 746 287 tweets for 47 stocks over 231 days (2017) | English | https://github.com/wuhuizhe/CHRNN |
| Headline [74] | News classification | 11 412 financial news headlines | English | https://www.kaggle.com/datasets/daittan/gold-commodity-news-and-dimensions |
| ConvFinQA [75] | Question answering | 3 892 dialogues with 14 115 question answering pairs | English | https://github.com/czyssrs/ConvFinQA |
| FLUE [34] | Multi-task evaluation | Aggregates FPB, FiQA, NER, Headline, and ConvFinQA | English | https://github.com/SALT-NLP/FLANG |
| FLARE [44] | Multi-task evaluation | Unified benchmark combining 9 + FinNLP datasets | English | https://github.com/The-FinAI/PIXIU |
| FinEval [63] | Multi-task evaluation | 8 351 question answering samples from exams, textbooks, and websites | Chinese | https://github.com/SUFE-AIFLM-Lab/FinEval |
| CFBenchmark [64] | Multi-task evaluation | 3 917 financial texts across 8 tasks for recognition, classification, and generation | Chinese | https://github.com/TongjiFinLab/CFGPT/tree/main/benchmark |
| FinanceIQ [65] | Multi-task evaluation | 7 173 multiple-choice questions across 10 topics | Chinese | https://github.com/Duxiaoman-DI/XuanYuan |
| FinBen [69] | Multi-task evaluation | 36 datasets covering 24 FinNLP tasks | English | https://github.com/The-FinAI/PIXIU |
| AlphaFin [70] | Multi-task evaluation | ~220 000 question answering samples (stock/news/report) with retrieval + CoT | English | https://github.com/AlphaFin-proj/AlphaFin |
| ICE-FLARE [51] | Cross-lingual task evaluation | 21 tasks covering question answering and classification | English | https://github.com/YY0649/ICE-PIXIU |
| CFinBench [66] | Multi-task evaluation | 99 100 mock exam questions across law, finance, and practice | Chinese | https://cfinbench.github.io/ |
| FinGEITje [72] | Cross-lingual task evaluation | 140 000 Instruction-tuning samples + Dutch question answering benchmark | Dutch | https://github.com/snoels/fingeit |
| CFLUE [68] | Multi-task evaluation | > 38 000 multiple-choice questions + 16 000 application tests | English Chinese | https://github.com/aliyun/cflue |
| $M^3$FinMeeting [71] | Cross-lingual task evaluation | Meeting dialogues in 11 sectors with summarization and question answering pairs | English Chinese Japanese | https://github.com/aliyun/qwen-dianjin |
| FLAME [67] | Multi-task evaluation | 16 000 exam questions + 100 tasks across 10 real-world scenarios | Chinese | https://github.com/FLAME-ruc/FLAME |

FPB: Financial PhraseBank; FiQA: financial question answering; ConvFinQA: conversational financial question answering; FLUE: financial language understanding evaluation; FLARE: financial language model assessment and risk evaluation; ICE-FLARE: InternLM-based financial language model assessment and risk evaluation; CFLUE: Chinese financial language understanding evaluation; FinEval: financial evaluation benchmark; CFBenchmark: Chinese financial benchmark 2022; CFinBench: Chinese financial benchmark 2023; FLAME: financial language model evaluation; NER: named entity recognition; FinanceIQ: financial intelligence quotient; FinBen: financial benchmark; AlphaFin: alpha finance; $M^3$FinMeeting: multimodal financial meeting; FinNLP: financial natural language-processing.

#### *2.3.1. Early stage: Task-specific and English-centric*

Initial FinNLP datasets were typically small-scale, focused on a single task, and exclusively in English. For example, the Financial PhraseBank (FPB) [59] provided sentence-level annotations for sentiment analysis; financial question answering (FiQA) dataset [60] introduced fine-grained annotations across microblogs and headlines for sentiment and question-answering (QA) tasks; and ACL18 stock movement prediction dataset [61], along with CIKM18 stock movement prediction dataset [62], collected tweets for stock movement prediction. These datasets enabled early FinLFMs such as FinBERT-19 [23], FLANG [34], and BloombergGPT [15] to demonstrate domain adaptation and transfer learning capabilities.

#### *2.3.2. Middle stage: Multi-task integration and language expansion*

The next wave of dataset development involved integrating multiple FinNLP tasks and expanding linguistic diversity. Financial language understanding evaluation (FLUE) dataset [34] and financial language model assessment and risk evaluation (FLARE) dataset [44] consolidated earlier benchmarks into unified evaluation suites, covering sentiment analysis, QA, named entity recognition, and forecasting. Parallel efforts in the Chinese financial domain resulted in datasets such as financial evaluation benchmark (FinEval) [63], Chinese financial benchmark 2022 (CFBenchmark) [64], financial intelligence quotient (FinanceIQ) dataset [65], Chinese financial benchmark 2023 (CFinBench) [66], and financial language model evaluation (FLAME) dataset [67], which simulate certification exams and textbook-style reasoning tasks. These multilingual, multi-task resources have significantly broadened the benchmarking landscape for non-English FinLFMs.

*2.3.3. Recent stage: Cross-lingual and real-world benchmarks*

More recent datasets push toward cross-lingual, instruction-tuned, and real-world financial understanding. Datasets such as InternLM-based financial language model assessment and risk evaluation (ICE-FLARE) [51], Chinese financial language understanding evaluation (CFLUE) [68], and financial benchmark (FinBen) dataset [69] offer multi-task evaluation across English and Chinese in retrieval, QA, and classification settings. The alpha finance (AlphaFin) dataset [70] integrates large-scale CoT-enhanced QA samples with document retrieval, targeting complex financial reasoning. Meanwhile, multimodal financial meeting ($M^3$FinMeeting) dataset [71] and Dutch financial dataset (FinGEITje) [72] introduce multilingual financial dialogues, meeting summaries, and Dutch financial QA tasks, emphasizing realistic, interactive applications. These resources mark a shift toward comprehensive, practical benchmarks that reflect the practical requirements of advanced FinLFMs.

*2.4. Summary*

FinLFMs have evolved rapidly from BERT-style encoders to GPT-style chatbots and reasoning-enhanced agents. This evolution has been driven by advances in updated and open-source base LFMs, scalable training pipelines, and increasingly diverse financial datasets. While FinLFMs now serve as the backbone for a wide range of FinNLP tasks, their limitations in handling temporal dynamics and multimodal content underscore the need for FMs that go beyond language. In the following sections, we turn our attention to FinTSFMs and FinVLFMs, which are intended to complement FinLFMs and build more holistic financial AI systems.

## 3. Financial time-series foundation models

This section introduces FinTSFMs, which are developed to analyze financial time-series data, including stock price, order book flow, and so forth. Similar to FinLFMs, we categorize FinTSFMs from three analytical perspectives: ① collection and category, ② training strategies, and ③ financial time-series datasets.

*3.1. Model classification*

While FinLFMs have rapidly matured over recent years, the exploration of FinTSFMs remains at an early stage. In this section, we provide a representative overview of existing FinTSFM efforts. Through an extensive literature search, we identify and summarize seven key models that exemplify the current landscape of FinTSFM research. As illustrated in Fig. 3, these models can be broadly categorized into two types: those trained from scratch solely on time-series data and those adapted from pre-trained LFMs using both time-series and related textual or structured inputs.

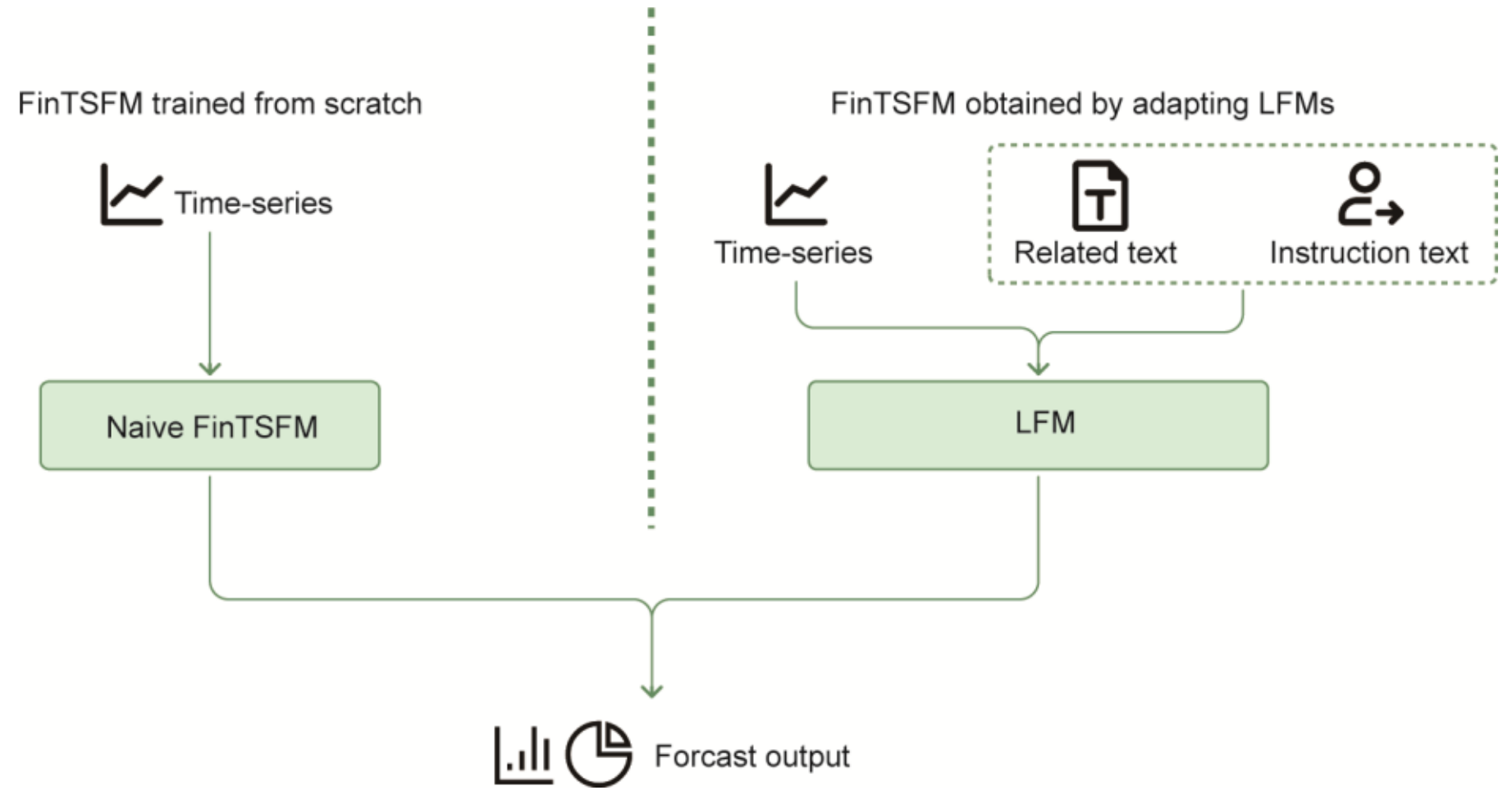

**Fig. 3.** Two categories of FinTSFMs: (left) models trained from scratch on time-series data and (right) models adapted from LFMs using both time-series and related textual data.

*3.1.1. Naive FinTSFMs trained from scratch*

Naive FinTSFMs are pre-trained from scratch on financial time-series data using transformer-based architectures. A representative example is MarketGPT [76], which uses a decoder-only transformer trained directly on National Association of Securities Dealers Automated Quotations (NASDAQ) order flow data. It is designed to act as an order-generation engine within discrete event simulations. Another milestone in this category is time-series foundation model (TimesFM) [77], a general-purpose time-series foundation model trained on real-world and synthetic time-series data in multiple domains including finance, transportation, and weather. Building on TimesFMs, several studies have adapted TimesFM for financial scenarios. For example, financial adaptation of TimesFM (Fin-TimesFM) [27] and financial-data-augmented TimesFM (FinDA-TimesFM) [78] extend TimesFM via continual pre-training on stock, currency, and crypto time-series data to improve domain specificity. Other recent studies [79,80] have explored how TimesFM can be adopted for value-at-risk (VaR) and realized volatility forecasting. However, a limitation of naive FinTSFMs is that they typically handle only numerical time-series data, lacking multimodal capabilities (e.g., joint text–time reasoning), which constrains their versatility in real-world financial scenarios.

### *3.1.2. FinTSFMs adapted from language models*

LLMs are inherently designed for sequential data and have demonstrated strong cross-domain generalization due to their pre-training on large-scale web corpora. Their adaptation to time-series tasks is motivated by two key factors: ① the efficacy of transformer-based models in long-range sequence modeling [81,82], and ② their ability to integrate multimodal knowledge, including numerical reasoning [83,84]. In this category, a time-series reprogramming LLM (Time-LLM) [85] introduces a reprogramming technique to convert time series into natural language-like prompts, which are then fed into a frozen LLM using the prompt-as-prefix strategy for time-series reasoning. A unified time-series model (UniTime) [86] extends this idea by integrating domain-specific instructions and multivariate time series as input to a finetuned GPT-2 model, achieving competitive results. Furthermore, SocioDojo [87] offers a unique non-training approach, leveraging the agentic capabilities of GPT-3.5 and GPT-4. It combines structured knowledge bases, Internet tools, and high-quality socioeconomic time series to evaluate AI agents on a financial time-series value prediction task called "hyperportfolio." Overall, FinTSFMs adapted from language models retain the powerful language understanding capabilities of LLMs while expanding into the time-series domain, making them more suitable for broad and practical financial applications.

## *3.2. Training method analysis*

We summarize the training strategies of representative FinTSFMs in Table 4 [27,76–78, 85–88]. These models can be grouped into three categories based on their adaptation paradigm: time-series pre-training, training-based LLM adaptation, and non-training methods.

**Table 4**

Training details of representative FinTSFMs.

| Dataset | Backbone | Method | Training data | Reference | Open source |
|---|---|---|---|---|---|
| MarketGPT | Transformer | Time-series pre-training | NASDAQ data | [76] | https://github.com/aaron-wheeler/MarketGPT |
| TimesFM | Transformer | Time-series pre-training | Google trends, wiki statistics, and synthetic data | [77] | https://huggingface.co/google/timesfm-1.0-200m |
| Fin-TimesFM | TimesFM | Time-series pre-training | Stocks, indices, foreign exchange, and crypto | [27] | https://github.com/pfnet-research/timesfm_fin |
| FinDA-TimesFM | TimesFM | Time-series pre-training | Stock price time-series | [78] | — |
| Kronos | Transformer | Time-series pre-training | Financial candlestick (*K*-line) sequences | [88] | https://github.com/shiyu-coder/Kronos |
| Time-LLM | GPT-2 | Training-based adaptation | Public datasets | [85] | https://github.com/KimMeen/Time-LLM |
| UniTime | GPT-2 | Training-based adaptation | Public datasets | [86] | https://github.com/liuxu77/UniTime |
| SocioDojo | GPT-3.5/4 | Non-training | — | [87] | https://github.com/chengjunyan1/SocioDojo |

### *3.2.1. Time-series pre-training*

This strategy involves pre-training FinTSFMs directly on time-series data, typically using autoregressive objectives analogous to those used in language modeling. However, due to the continuous and multivariate nature of time-series inputs, these models often incorporate patching techniques [89] to divide input sequences into fixed-length segments. A representative example is TimesFM [77], along with its financial-domain adaptations such as Fin-TimesFM [27] and FinDA-TimesFM [78]. These models adopt patch-wise encoding to capture temporal dependencies across multiple correlated variables. Another notable design is MarketGPT [76], a decoder-only transformer that tokenizes raw order-message data from electronic trading systems into discrete event sequences. This transformation enables autoregressive learning over market microstructure events, diverging from conventional numerical patching approaches and representing a unique trajectory for financial sequence modeling. Most recently, Kronos [88] introduces a language-inspired formulation that represents financial candlestick (*K*-line) time-series as a symbolic vocabulary. By embedding open–high–low–close–volume–amount (OHLCVA) patterns into discrete tokens, Kronos treats market dynamics as a structured "language," allowing large-scale pre-training over extended price trajectories. This symbolic design highlights interpretable temporal motifs and demonstrates strong performance in tasks such as market regime detection and volatility forecasting.

### *3.2.2. Training-based LLM adaptation*

This approach adapts general-purpose language models to time-series tasks via additional training. A key distinction among existing methods lies in whether the LLM's parameters are updated. For instance, Time-LLM [85] preserves all frozen parameters of the backbone LLM while introducing an adaptation layer that reprograms time-series segments into prompt-like representations for reasoning. In contrast, UniTime [86], a fine-tuned GPT-2 variant, updates the LLM parameters directly using multivariate time-series data and domain-specific instruction tuning, thereby enabling it to model temporal dependencies natively. The contrast between these two approaches reflects the absence of a standardized adaptation pipeline for FinTSFMs—unlike FinLFMs, which increasingly follow a pre-training/SFT/alignment paradigm. This divergence suggests that training strategies for FinTSFMs remain exploratory and ripe for future formalization and innovation.

**Table 5**

Details of common financial time-series-related datasets.

| Dataset | Scale | Task(s) | Reference | Open source |
|---|---|---|---|---|
| Google Stock Prices | Daily, 2012–2017 | Stock price prediction | [90] | https://www.kaggle.com/datasets/vaibhavsxn/google-stock-prices-training-and-test-data |
| S&P 500 Index | Daily, 1927–2020 | Stock price prediction | [91] | https://www.kaggle.com/datasets/henryhan117/sp-500-historical-data |
| Exchange Rate (8 currencies) | Daily, 1990–2016 | Exchange rate prediction | [92] | https://github.com/laiguokun/multivariate-time-series-data |
| Bitcoin Prices | Daily, 2010–2020 | Cryptocurrency trend forecasting | [93] | https://www.kaggle.com/datasets/soham1024/bitcoin-time-series-data-till-02082020 |
| FNSPID (news + prices) | 1999–2023, 29.7 million prices, and 15.7 million news | Integrated market analysis and price forecasting | [94] | https://huggingface.co/datasets/Zihan1004/FNSPID |
| FinTSB | Comprises 20 datasets, each with 300 stocks over 250 days, grouped into 4 distinct market patterns: uptrend, downtrend, volatility, and black swan | Stock movement forecasting under realistic constraints | [95] | https://github.com/TongjiFinLab/FinTSBenchmark |

FNSPID: financial news and stock price integration dataset; FinTSB: financial time-series benchmark.

### *3.2.3. Non-training methods*

SocioDojo [87] exemplifies an alternative to training-intensive methods by leveraging LLMs as reasoning agents without explicit time-series fine-tuning. Built upon GPT-3.5 or GPT-4, SocioDojo equips LLMs with access to external tools, including Internet-based application programming interfaces (APIs), knowledge bases, and analytical modules. These agents interact with structured time-series inputs through multi-step planning and decision-making workflows, enabling tasks such as hyperportfolio construction and dynamic forecasting. This approach underscores the potential of agentic LLM systems to reason over time-series data without extensive retraining, paving the way for lightweight and tool-augmented FinTSFMs.

## *3.3. Financial time-series-related datasets*

Commonly used datasets for financial time-series modeling are summarized in Table 5 [90–95]. These datasets vary in scale, format, and task coverage. For instance, the Google Stock Prices dataset [90] provides daily stock data from 2012 to 2017 and is widely used for basic market trend prediction. The S&P 500 Index dataset [91] offers a long historical span (1927–2020), supporting index-level forecasting and macroeconomic analysis. The Exchange Rate dataset [92], which covers the daily rates of eight currencies from 1990 to 2016, is frequently used to evaluate long-horizon forecasting models. The Bitcoin Prices dataset [93] tracks cryptocurrency price fluctuations and serves as a benchmark for volatile financial assets. For multimodal or multi-source financial applications, financial news and stock price integration dataset (FNSPID) [94] is a large-scale dataset integrating over 29.7 million price records and 15.7 million news headlines across more than 4000 listed companies, enabling the joint modeling of time series, event impacts, and sentiment effects. Most recently, financial time-series benchmark (FinTSB) [95] has been introduced as a standardized benchmark for financial time-series forecasting. It addresses key limitations of prior datasets, including limited pattern diversity, inconsistent evaluation, and unrealistic trading assumptions. FinTSB consists of 20 datasets across four market regimes (uptrend, downtrend, volatility, and black swan), each containing 300 stocks over 250 days. It offers unified metrics (ranking, portfolio, and error), simulates real-world constraints (e.g., transaction fees), and provides detailed sequence characteristics (e.g., non-stationarity and forecastability).

Despite recent progress, the development of financial time-series datasets remains in its early stages. Most existing resources are constrained by limited scope or short temporal coverage, hindering the comprehensive evaluation of long-context modeling, reasoning, and generalization. Continued efforts are needed to build larger, more diverse, and practically grounded benchmarks to support future advancements of FinTSFMs.

## *3.4. Summary*

FinTSFMs are still in an early yet rapidly evolving stage. Current efforts range from scratch pre-training to LLM adaptation and non-training approaches, but a standardized pipeline like that of FinLFMs has not emerged. While new datasets such as FNSPID [94] and FinTSB [95] broaden the evaluation scope, most resources remain limited in scale or realism, hindering the assessment of long-horizon and multimodal capabilities. Recent work, such as that by Marconi [96], has demonstrated that pre-training time-series foundation models (TSFMs) increases sample efficiency and transferability across tasks including treasury yield, foreign exchange volatility, and equity spread forecasting. These findings highlight both the promise of FinTSFMs and the need for domain-specific pre-training, task-oriented refinement, and more robust benchmarks.

## 4. Financial visual-language foundation models

This section focuses on FinVLFMs, which aim to jointly process financial visual information (e.g., line charts, candlestick diagrams, and scanned reports) and associated textual context to support tasks such as visual question answering (VQA), document parsing, and multimodal reasoning in financial domains. A representative case is the FOMC press meeting, where monetary policy is communicated not only through textual statements and Q&A transcripts but also through visual artifacts such as the interest rate "dot plot" and projection charts. These meetings exemplify the need for FinVLFMs to integrate structured visual signals with natural language context in order to extract insights and support downstream tasks such as policy impact assessment and market reaction forecasting. Like FinTSFMs, FinVLFMs are in early developmental stages, with only a few representative works following relatively consistent design patterns. Our analysis introduces FinVLFMs through three lenses: ① common architectural components, ② training methodologies, and ③ specialized datasets that combine financial visuals (e.g., charts and report images) with textual content (e.g., news articles and regulatory filings). In what follows, we detail each component individually and discuss how FinVLFMs may benefit from techniques developed in general-purpose VLMs.

### *4.1. Common FinVLFM architecture components*

As shown in Fig. 4, most FinVLFMs follow a three-module architecture:

- **Vision encoder.** This module involves a pre-trained image encoder (e.g., contrastive language–image pre-training (CLIP) model [97]) that transforms financial visuals into dense feature embeddings.
- **Vision projector.** This module involves a lightweight adapter (e.g., a multi-layer perceptron (MLP)) that aligns visual embeddings with the latent space of the LLM.
- **Base LLM.** This module involves a powerful LFM that integrates visual and textual inputs for downstream financial tasks.

In what follows, we detail each component individually and discuss how FinVLFMs may benefit from techniques developed in general-purpose VLMs.

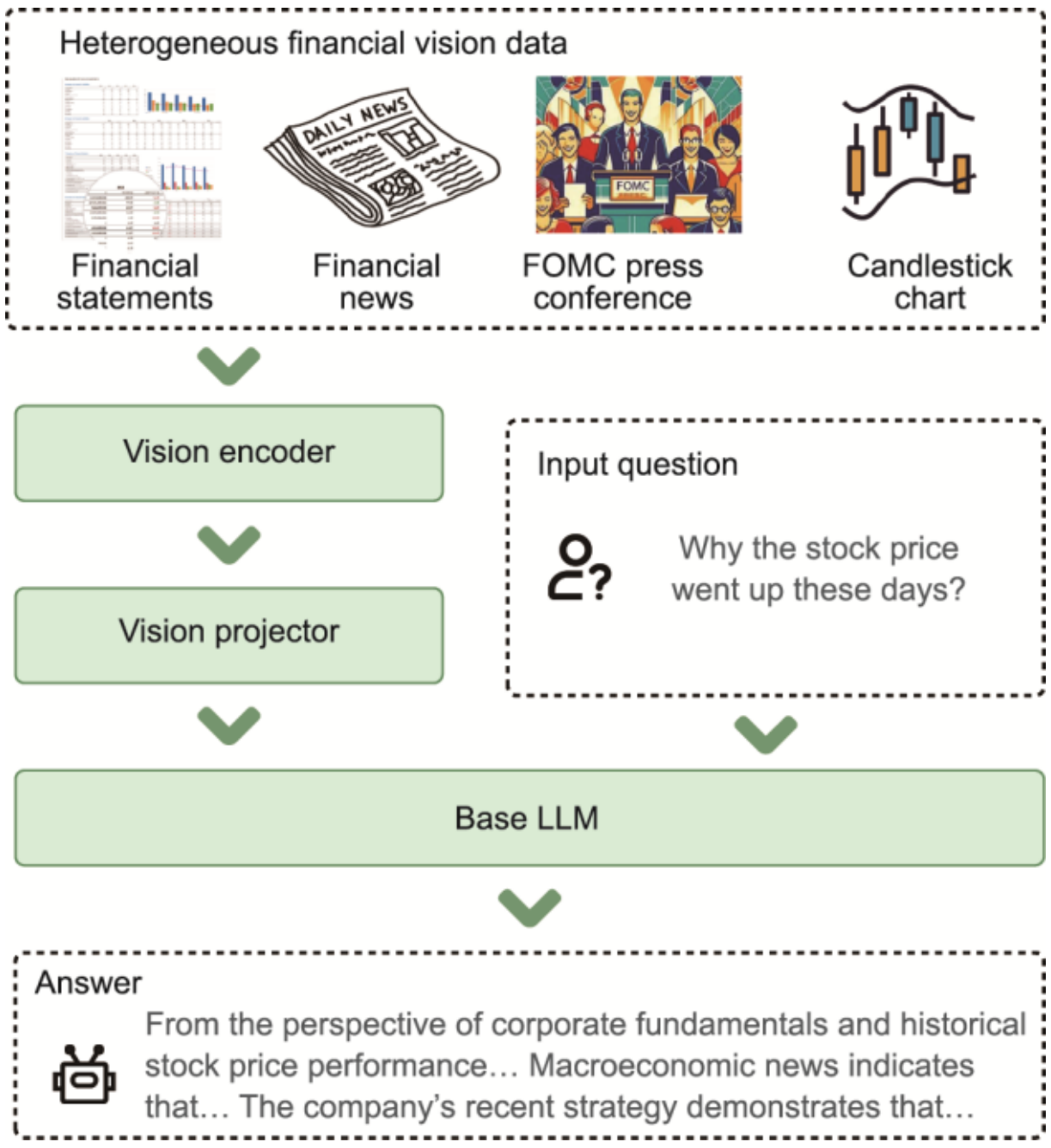

**Fig. 4.** Current FinVLFMs share three common components: a vision encoder, vision projector, and base LLM.

#### *4.1.1. Vision encoder*

The vision encoder converts diverse financial visual inputs (e.g., line charts, bar graphs, and tables) into embedding representations. Most existing FinVLFMs (e.g., financial large language and vision assistant (FinLLaVA) [19] andmultimodal financial LLM built on Mistral (FinTral) [20]) employ CLIP or other general-purpose encoders. While convenient, this design overlooks the domain-specific visual patterns in finance. For example, line charts often require fine-grained resolution, and tabular data demands cell-level parsing—tasks not well-captured by generic visual encoders. Recent progress in general vision-language models (VLMs) offers promising pathways. Bootstrapping language-image pre-training (BLIP) [98] improves image–text alignment through bootstrapped contrastive learning, while Large multimodal model perceiving any aspect ratio and high-resolution images (LLaVA-UHD) [99] enhances detail resolution using adaptive image patching. These methods can be directly adapted to improve the visual understanding capabilities of FinVLFMs.

#### *4.1.2. Vision projector*

The vision projector aligns the visual embedding space with the LLM's token space, acting as the critical modality adapter. In current FinVLFMs, this layer is usually implemented using simple MLPs or linear projections (e.g., FinLLaVA [19] uses a two-layer MLP). While sufficient for coarse-grained tasks, such basic aligners may struggle with complex visual semantics in financial reports. In contrast, general-domain VLMs increasingly employ cross-attention mechanisms [100,101], gating layers, or dynamic adapters that better preserve inter-modal dependencies. Applying such techniques to FinVLFMs can improve interpretability and context-aware reasoning, especially in tasks involving tabular-cell and chart-element mapping.

#### *4.1.3. Base LLM*

The base LLM decodes aligned multimodal representations and performs financial reasoning. The early financial vision GPT (FinVis-GPT) used Vicuna [102] as the backbone decoder. However, domain mismatch limits financial fluency. To address this issue, FinLLaVA [19] introduced FinLLaMA, an LLaMA-based model further pretrained on financial corpora. Similarly, FinTral [20] leverages Mistral-7B for its superior numerical handling. These adaptations underscore the importance of integrating a FinLFM in FinVLFM pipelines, particularly for tasks such as earnings call summarization, portfolio optimization, or risk-disclosure analysis.

### *4.2. Training method analysis*

Table 6 [18–20] summarizes representative FinVLFMs, including their model backbones, parameter scales, training methods, and dataset sizes. Despite architectural diversity, these models consistently follow a two-stage training paradigm: modal alignment pre-training to align cross-modal embeddings, and SFT to improve finance-specific reasoning.

**Table 6**

Training details of FinVLFMs.

| Model | Backbone | Parameter | Training method | Training dataset size | Reference | Open source |
|---|---|---|---|---|---|---|
| FinVis-GPT | Vicuna | — | Modal alignment PT and SFT | 300 000 VQA pairs | [18] | https://github.com/wwwadx/FinVis-GPT |
| FinTral | Mistral | 7 billion | Modal alignment PT and SFT | 1.86 million VQA pairs | [20] | https://github.com/UBC-NLP/fintral |
| FinLLaVA | FinLLaMA | 8 billion | Modal alignment PT and SFT | 1.43 million VQA pairs | [19] | https://www.thefin.ai/model/open-finllms |

#### *4.2.1. Modal alignment pre-training*

The first stage focuses on aligning the visual encoder and the language model within a shared semantic space. This is typically achieved by freezing the backbone components (i.e., the vision encoder and the LLM) and training the visual projection layers. All three FinVLFMs adopt this alignment-first approach, but they differ in their alignment data construction strategies. FinVis-GPT [18] uses historical Chinese A-share market data to generate candlestick charts with technical indicators, supplemented by textual explanations generated by ChatGPT. FinTral [20] and FinLLaVA [19] integrate large-scale generic visual datasets with domain-specific financial data derived from stock price charts.

#### *4.2.2. Supervised fine-tuning*

After modal alignment, the models undergo supervised fine-tuning to improve response quality and financial task alignment. Unlike the alignment stage, this phase often unfreezes both the LLM and the projection layer, enabling full parameter optimization. FinVLFMs differ in the granularity and content of their instruction datasets. Regardless of the data source, the goal of this phase is to teach the model how to follow financial-specific prompts, interpret visual evidence, and generate coherent, grounded responses.

### *4.3. Financial visual-language-related datasets*

Table 7 [19,20,103–113] summarizes commonly used financial visual-language-related datasets. These datasets cover a variety of modalities, such as charts, tables, and documents. In the context of FinVLFMs, however, all non-text modalities—including tables—are typically rendered as images and processed through a unified visual encoder. This design simplifies the system pipeline and provides a consistent input format, which is beneficial for downstream financial applications.

**Table 7**

Details of common financial (VQA datasets.

| Benchmark | Data type | Task | Size | Reference | Open source |
|---|---|---|---|---|---|
| Australian Credit | Table (credit records) | Credit scoring | 690 | [103] | https://archive.ics.uci.edu/dataset/143/statlog+australian+credit+approval |
| German Credit | Table (credit records) | Credit scoring | 1 000 | [104] | https://archive.ics.uci.edu/dataset/144/statlog+german+credit+data |
| TAT-QA | Text and table | Numerical VQA | 19 309 | [105] | https://nextplusplus.github.io/TAT-QA/ |
| FinQA | Text and table (earnings) | Financial VQA | 8 281 | [106] | https://github.com/czyssrs/FinQA |
| ChartQA | Chart | Chart VQA | 2 500 | [108] | https://github.com/vis-nlp/ChartQA |
| ConvFinQA | Text and table (earnings) | Financial domain VQA | 3 892 | [107] | ttps://github.com/czyssrs/ConvFinQA |
| MMMU Business | Charts, diagrams, and tables | Business VQA | 1 428 | [110] | https://mmmu-benchmark.github.io/ |
| FinVQA | Chart (finance and stock market) | Chart VQA | 1 025 | [20] | https://github.com/UBC-NLP/fintral |
| ChartBench | Chart | Reasoning over charts | 350 | [19] | https://chartbench.github.io/ |
| TableBench | Table | Reasoning over tables | 450 | [19] | https://tablebench.github.io/ |
| MME-Finance | Charts, tables, and documents | Financial domain VQA | 1 171 | [109] | https://hithink-research.github.io/MME-Finance/ |
| FCMR | Text, table, image, and chart | Multi-hop VQA | 2 199 | [111] | — |
| FAMMA | Text, charts, and tables | Financial domain VQA | 1 758 | [112] | https://famma-bench.github.io/famma/ |
| FinMME | Text, charts, and tables | Financial domain VQA | 11 099 | [113] | https://huggingface.co/datasets/luojunyu/FinMME |

TAT-QA: tabular and textual dataset for question answering; MMMU: massive multi-discipline multimodal understanding; ChartQA: chart question answering; FCMR: financial cross-modal retrieval; FAMMA: financial assistant multimodal multi-task assessment; FinMME: financial multimodal evaluation; FinVQA: financial visual question answering; ChartBench: chart understanding and reasoning benchmark; TableBench: table understanding and reasoning benchmark; MME-Finance: multimodal evaluation finance subset.

As a result, most tasks across these datasets can be consistently categorized as VQA, regardless of the original modality. For example, datasets such as Australian Credit and German Credit [103,104] provide table-based VQA tasks targeting credit scoring scenarios. Tabular and textual dataset for question answering (TAT-QA) [105], FinQA [106], and conversational financial question answering (ConvFinQA) [107] focus on numerical VQA tasks grounded in financial earnings reports, often involving complex multi-step reasoning over visualized tabular data.

Chart-oriented datasets such as chart question answering (ChartQA) [108], financial visual question answering (FinVQA) [20], and chart understanding and reasoning benchmark (ChartBench) [19] evaluate a model's ability to interpret line charts, bar plots, and other financial graphics for answering quantitative or descriptive questions. Multimodal evaluation finance subset (MME-Finance) [109] further expands this by integrating multiple financial visual formats (i.e., charts, tables, and documents) into a unified evaluation suite for diverse VQA tasks.

In addition, massive multi-discipline multimodal understanding (MMMU) Business [110], financial cross-modal retrieval (FCMR) [111], and financial assistant multimodal multi-task assessment (FAMMA) [112] propose more complex, multimodal or multi-hop financial VQA settings. These benchmarks introduce questions requiring reasoning over multiple types of inputs, integration of domain knowledge, and higher-level abstraction beyond single-modal analysis. Most recently, financial multimodal evaluation (FinMME) [113] has collected over 10 000 high-quality VQAs with detailed annotation, aiming to establish a new benchmark for FinVLFMs.

Despite the diversity of content and design, most current datasets remain limited in scale, typically offering only hundreds of to a few thousand VQA pairs. Thus, they are best suited for evaluating early-stage FinVLFMs within a multidimensional evaluation ecosystem and motivating the development of more robust, large-scale, and finance-specialized multimodal benchmarks. In particular, while these datasets are valuable for performance benchmarking, they are not sufficient to support the pre-training or large-scale instruction tuning of FinVLFMs. Future research will require the construction of significantly larger, more diverse, and task-rich financial visual-language datasets encompassing real-world financial charts, tables, filings, and reports to unlock the full potential of FinVLFMs.

### *4.4. Summary*

FinVLFMs represent a promising yet nascent direction in financial AI, enabling multimodal understanding across charts, documents, and textual narratives. Current models converge on a common three-component architecture and two-stage training paradigm, yet remain constrained by limited data scale, domain-specific visual encoding, and instruction diversity. Recent work such as vision-language inference for training-free stock time-series analysis (VISTA) [114] has shown that vision-language inference can directly support stock forecasting in a training-free manner, while Quant 4.0 [115] positions multimodal and knowledge-driven AI as the next frontier of quantitative investment. These advances suggest that FinVLFMs hold promise not only as technical enablers but also as integral components of future financial engineering pipelines.

## **5. FFM-based financial applications**

In this section, we review representative applications involving FFMs. A survey of the literature revealed that many existing studies still rely on general-purpose FMs (e.g., GPT-4 and Gemini) to explore the feasibility and potential of FMs in various financial tasks. However, several emerging works have begun to employ domain-specific FFMs (e.g., ICE-INTENT, TimesFM, and CFGPT) that demonstrate clear advantages in bilingual understanding, compliance, and time-series prediction. Therefore, we have selected 11 representative application-oriented studies, encompassing both general-purpose and domain-specialized FMs, to illustrate the current landscape of FFM-based financial applications. We categorize these applications into four types: ① financial knowledge extraction, ② market prediction, ③ trading and financial decision-making, and ④ agent-based financial simulation. Table 8 [51,79,116–126] summarizes representative studies within each category, with domain-specific FFMs highlighted in bold.

**Table 8**

An overview of FFM-based applications with representative works.

| Task | Reference | Involved FM(s) | Description |
|---|---|---|---|
| Financial knowledge extraction | | | |
| Bilingual financial text understanding | [51] | **ICE-INTENT, DISC-FinLLM, CFGPT,** and GPT-4 | ICE-INTENT supports 18 bilingual tasks and surpasses GPT-4 and other FinLFMs in bilingual financial understanding |
| Financial relation extraction | [116] | GPT-4, pathways language model 2 （PaLM2） and mosaic pretrained transformer (MPT) | Zero-shot annotation of financial relations for efficient and reliable dataset creation |
| Financial table parsing | [117] | GPT-4 | Accurate extraction of values from annual report tables |
| Market prediction | | | |
| VaR forecasting | [79] | **TimesFM** | Evaluates TimesFM for left-tail VaR prediction on S&P 100 Index, showing superior performance over traditional models |
| Stock ranking from multi-source signals | [118] | GPT-4 | Outperforms S&P 100 Index with CoT investment signals |
| Sentiment-driven trading | [119] | GPT-4 | Improves timing by filtering investor sentiment |
| Signal enhancement qualitative portfolio insight generation | [120] | GPT-4 | Assists stock selection and diversification with GPT-driven analysis |
| Trading and financial decision | | | |
| Retrieval-augmented-generation-based financial assistant system | [121] | **CFGPT** | Introduces a retrieval-augmented FinLFM with a hybrid knowledge base, and compliance-aware checkers for tasks such as question and answering, investment advising, and risk control |
| Trading with memory and personality | [122] | GPT-4 | Introduces a layered memory architecture and persona-driven design to improve trading decision-making |
| Explainable factor extraction | [123] | GPT-4 | Uses customized prompting to extract interpretable market-moving factors from news |
| Portfolio diversification support | [120] | GPT-4 | Improves portfolio diversification and returns when combined with quant optimization |
| Agent-based financial simulation | | | |
| Role-specialized trading | [124] | GPT-4o, OpenAI-o1 | Multi-role AI agents collaborate to trade and manage risk |
| Simulation market theory Testing via agent simulation | [125] | GPT-4 | Multiple agents simulate realistic market to test financial theories |
| Investor behavior emulation | [126] | GPT-3.5, Gemini | Multiple agents simulate different trading behaviors to real news and price data |

Most studies adopt general-purpose FMs for initial exploration; domain-specific FFMs are highlighted in bold.

### *5.1. Financial knowledge extraction*

A fundamental application of FFMs is the transformation of unstructured financial text, tables, and reports into structured knowledge representations that directly support downstream analysis, prediction, and decision-making. Rather than serving merely as a preprocessing step, financial knowledge extraction should be viewed as an end-to-end application that enables systematic knowledge management in finance. Existing studies emphasize this potential across diverse scenarios: Hu et al. [51] showed that the domain-specialized ICE-INTENT model outperforms GPT-4 and other FinLFMs in bilingual financial understanding; Aguda et al. [116] employed GPT-4 in a zero-shot setting for relation extraction by generating tuples such as <organization, date>; and recent work on annual report analysis has demonstrated how LLMs can parse financial tables to extract key corporate items [117]. Together, these efforts demonstrate that FFMs function not only as technical enablers but also as practical tools for financial knowledge extraction, converting raw documents into structured knowledge that forms the basis for applications such as market prediction, compliance monitoring, and trading decision support.

### *5.2. Market prediction*

FFMs have also been increasingly applied to market forecasting tasks, such as predicting asset risk, market sentiment, and timing signals from financial text and time-series data. Goel et al. [79] were recently among the first to apply a TimesFM to VaR forecasting, estimating left-tail return quantiles for assets in the S&P 100 Index. While such efforts highlight the emerging role of FinTSFMs, most current studies still rely on general-purpose LLMs to explore the feasibility of market prediction tasks in financial contexts. Fatouros et al. [118] developed MarketSenseAI, prompting GPT-4 with multi-source signals including fundamentals, macroeconomic indicators, and news headlines to generate investment rationales that outperform the S&P 100 Index. Similarly, Chen et al. [119] proposed a multimodal framework in which GPT-4 evaluates investor forum comments to enhance sentiment-driven stock timing, improving profitability and win rates in backtests. Ko and Lee [120] investigated the role GPT-4 can play in portfolio analysis, finding that it offers useful qualitative insights for stock selection and diversification when used alongside quantitative methods. These studies have collectively shown that, while general-purpose LLMs dominate current explorations, FFMs—particularly FinTSFMs—are beginning to demonstrate unique advantages, paving the way for more specialized solutions in financial prediction.

### *5.3. Trading and financial decision-making*

Beyond forecasting, FFMs are increasingly deployed in end-to-end financial decision-making pipelines, encompassing trading strategy development, investment advising, and risk control. For example, Li et al. [121] proposed a Chinese financial assistant named as RA-CFGPT, built upon a finetuned FinLLM, a hybrid knowledge base under a retrieval-augmented-generation framework, and compliance-aware checkers. The system demonstrates robust performance across tasks such as question answering, investment advising, and risk assessment. Other studies have explored the use of general LLMs for trading and financial decision-making. Here, a representative work is FinMem, proposed by Yu et al. [122]; it enhances LLM-based trading agents with a layered memory system and persona-driven architecture. By simulating short-, mid-, and long-term memory along with distinct investor styles (e.g., aggressive or conservative), FinMem achieves superior backtest performance compared with both deep reinforcement learning and vanilla LLM agents. Complementing this, Wang et al. [123] introduced LLMFactor, a sequential prompting framework that extracts interpretable alpha signals from financial news aligned with market dynamics, improving both predictive performance and interpretability. Finally, Ko and Lee [120] demonstrated that GPT-4-guided portfolio construction can yield improved risk-adjusted returns, especially when combined with classical quantitative optimization. Collectively, these works illustrate the diverse ways in which FFMs can be adapted for high-stakes decision-making under uncertainty.

### *5.4. Agent-based financial simulation*

Agent-based financial simulation is a rapidly emerging application area for FFMs, in which multiple AI-driven agents interact to emulate market dynamics and investor behaviors. Unlike generic multi-agent frameworks, this line of work explicitly targets financial contexts such as trading, risk management, and market microstructure analysis. For example, Xiao et al. [124] simulated realistic market interactions with agents assigned to trading-related roles (e.g., trader and risk officer), achieving coordinated decision-making through structured prompts. Lopez-Lira [125] simulated price formation and liquidity using GPT-4-driven market-making agents, offering new ways to test economic hypotheses. Similarly, StockAgent [126] emulates investor behavior in reaction to market news using multi-agent systems powered by GPT-3.5 and Gemini. While most current systems still rely on general-purpose LLMs, their methods can be naturally extended to FFMs. Incorporating domain-specific FFMs could increase realism, compliance alignment, and predictive fidelity in future simulations. Notably, several existing research groups [124,126] have explicitly listed the training of task-specific FFMs as part of their future work, suggesting that agents specialized for trading, risk management, or investor-behavior modeling may provide more accurate and domain-aligned outcomes. This indicates a clear trajectory from proof-of-concept multi-agent simulations toward FFM-driven agent architectures that embed financial expertise at the task level.

### *5.5. Summary*

Current FFM-based financial applications mainly rely on general-purpose FMs to test feasibility in knowledge extraction, market prediction, trading, and simulation. These studies demonstrate the exploratory potential of FMs in finance but often lack domain specificity for high-stakes use cases. However, emerging domain-specific FFMs (e.g., ICE-INTENT [51], CFGPT [52], and DISC-FinLLM [57]) are beginning to demonstrate advantages in financial understanding, compliance, and structured reasoning. These developments suggest a clear trajectory toward the integration of specialized FFMs, which are expected to improve accuracy, interpretability, and robustness in high-stakes financial environments.

## 6. Challenges and opportunities

Despite the rapid progress in developing and applying FFMs, their widespread adoption still presents significant obstacles arising not only from the limitations of current algorithmic designs and computational infrastructure but also—crucially—from the unique characteristics of financial data and tasks. Unlike general AI domains, financial engineering presents distinctive requirements such as regulatory compliance, market non-stationarity, structured multimodal data (e.g., candlestick charts, balance sheets, order flows), and latency-critical decision-making. Understanding and addressing these bottlenecks is essential for advancing FFMs from proof of concept to practical deployment. To provide a structured perspective, we categorize the key challenges and corresponding research opportunities into three dimensions: data, algorithm, and computing infrastructure. Each dimension reflects a crucial aspect of the FFM development pipeline, from data acquisition to system-level deployment. In what follows, we highlight critical limitations, emerging solutions, and representative works that pave the way for future progress.

### *6.1. Data*

#### *6.1.1. Scarcity of large-scale multimodal financial datasets*

As discussed in Sections 2–4, current FFMs, especially FinVLFMs and FinTSFMs, are constrained by the lack of large-scale, high-quality multimodal datasets (e.g., chart–text or table–report pairs). Unlike natural images or videos in general AI, financial multimodal data is highly structured, domain-specific, and often requires precise numerical interpretation. Researchers have highlighted this scarcity and the underdeveloped multimodal capabilities of FFMs [19], which limit these models' real-world applicability. However, acquiring such data from financial environments is costly and heavily regulated. An alternative strategy is data synthesis using advanced LLMs, which is emerging as a promising approach for constructing financial benchmarks. For example, He et al. [127] generated synthetic chart reasoning datasets by programmatically producing candlestick plots paired with high-quality QA items. Recent works such as FinTral [20] and FinLLaVA [19] also attempt to build chart-based QA datasets tailored for financial contexts. The automatic generation of accurate, diverse, and finance-specific multimodal data remains a key research direction.

#### *6.1.2. Privacy and regulatory constraints*

Many valuable financial datasets, such as internal bank reports, client transactions, and proprietary trading strategies, are confidential and inaccessible to the public due to privacy regulations (e.g., general data protection regulation (GDPR) [128], markets in financial instruments directive II (MiFID II) [129], and Basel III [130]) and competitive concerns. This severely limits data sharing and aggregation for FFM training, making the challenge far more pronounced than in other domains. A promising solution lies in federated learning, which allows multiple institutions (e.g., banks and hedge funds) to collaboratively train models without sharing raw data. Several works have already explored federated training paradigms for general-domain LLMs that preserve data privacy [131,132]. Domain-specific efforts such as the differential-privacy protection financial large language model (DPFinLLM) [133] offer a concrete attempt at the privacy-preserving fine-tuning of FinLFMs by combining low-rank adaptation with a differential privacy strategy. Recent models such as XuanYuan3 [49] and Baichuan4-Finance [54] incorporate compliance-focused alignment using reinforcement learning, explicitly penalizing responses that violate financial regulations. These approaches offer valuable technical foundations for privacy-preserving and regulation-aware FFM development in finance.

### *6.2. Algorithm*

#### *6.2.1. Hallucination and factual inconsistency*

FMs are prone to hallucination—that is, generating incorrect or fabricated information. In financial contexts, such errors can have severe consequences, particularly when models produce false statements regarding earnings, asset risks, or regulatory events. Even state-of-the-art systems such as DeepSeek-R1 have been shown to generate unverified or inconsistent financial claims [9]. Addressing this issue requires the integration of FFMs with structured knowledge sources. For instance, Jiang et al. [134] proposed a multimodal retrieval-augmented generation (RAG) system that combines a vector database and a financial knowledge graph. By converting visual data into text and retrieving structured facts across modalities, the system supports LLM reasoning grounded in reliable information. Such integration between FFMs and knowledge engineering offers a promising path toward trustworthy financial applications.

#### *6.2.2. Lookahead bias and market non-stationarity*

FFMs are typically trained on large corpora that may contain future information relative to backtest periods, introducing lookahead bias. This bias leads to inflated performance during model evaluation because the model has inadvertently learned about future events. For example, an FM evaluated on data from 2020 may have been trained on news published after 2020, violating the temporal integrity of the backtest [135]. To mitigate this problem, careful curation of temporally consistent datasets is essential. One example is TimeMachineGPT [136], in which models are trained exclusively on data constrained to specific time horizons to prevent future leakage. Developing temporally aware training and evaluation protocols remains a critical direction for FFMs. Beyond lookahead bias, financial time-series data is inherently non-stationary due to regime shifts, volatility clustering, and abrupt market shocks—properties rarely encountered in general natural language-processing (NLP) tasks. Recent models such as Fin-TimesFM [27], MarketGPT [76], and Kronos [88] attempt to address this issue through continual pre-training on streaming market data or by directly modeling order-book dynamics. Nevertheless, developing robust methods to adapt to structural breaks and rapidly changing financial environments remains an open research frontier.

### *6.3. Computing infrastructure*

#### *6.3.1. High training and deployment costs*

Training or fine-tuning FFMs with tens of billions of parameters is extremely resource-intensive, often requiring thousands of graphics processing units (GPUs), extensive memory, and weeks of computation time. This presents a significant financial and infrastructural barrier, especially for academic institutions and small enterprises. For example, BloombergGPT [15] was trained using approximately 1.3 million GPU hours on NVIDIA A100s, an effort estimated to cost 1–2 million USD. In addition to training costs, deploying large-scale FFMs poses substantial challenges. Inference with large models typically incurs high latency, substantial energy consumption, and complex infrastructure requirements, particularly in real-time or low-latency financial applications such as trading, risk monitoring, or customer interaction systems. In response, research has begun to explore lightweight alternatives such as distilled, small, yet powerful FFMs derived from general-purpose LLMs [9], which aim to retain financial domain capabilities while significantly reducing computational demands. Beyond simple model compression, an emerging paradigm advocates for a collaborative system that integrates large and small models. Researchers [137,138] have discussed hybrid systems in which large LLMs are used to provide domain knowledge, supervision, or feature augmentation, while lightweight models handle privacy-sensitive or latency-critical tasks on the edge. For example, RA-CFGPT [121] integrates a small seven billion self-trained Chinese financial expert model with GPT-4 and LLaMA into a chatbot system to balance latency and trustworthiness. This collaborative architecture is particularly attractive for financial scenarios with strict regulatory or cost constraints, suggesting that future FFMs may evolve into modular ecosystems of large and small models working in synergy.

## 7. Conclusions

FFMs are reshaping the landscape of financial engineering by enabling scalable, adaptable, and multimodal intelligence across tasks such as analysis, forecasting, and decision-making. In this survey, we systematically reviewed the architectural designs, training methodologies, benchmark datasets, and practical applications of FFMs, categorizing them into three main types: FinLFMs, FinTSFMs, and FinVLFMs.

Despite notable progress, FFMs still present a range of open challenges, from data availability and temporal consistency to model trustworthiness, efficiency, and domain adaptation. Tackling these issues will require advances across multiple fronts, including high-quality dataset construction, multimodal integration, efficient training and deployment strategies, and deeper alignment with financial knowledge and regulatory requirements. We hope this work provides a useful foundation for further research and encourages the responsible development of FFMs in real-world financial systems.

## Data availability

An updated collection of FFM-related publications and resources are accessible via the following link: https://github.com/FinFM/Awesome-FinFMs.